\documentclass[onecolumn,12pt]{article}

\usepackage{graphicx}
\usepackage{subfig}
\usepackage{amssymb,amsmath}
\usepackage{authblk}
\usepackage{cite}
\usepackage{setspace}

\title{Carrier multiplication between interacting nanocrystals for fostering silicon-based photovoltaics}
\author[1]{Marco Govoni\thanks{marco.govoni@unimore.it}}
\author[2]{Ivan Marri\thanks{ivan.marri@unimore.it}}
\author[2,3]{Stefano Ossicini}
\affil[1]{Department of Physics, University of Modena and Reggio Emilia, via Campi 213/A, 41125 Modena, Italy}
\affil[2]{Department of Science and Methods for Engineering (DISMI), via Amendola 2 Pad. Morselli, 42122 Reggio Emilia, Italy}
\affil[3]{CNR-NANO Research Center S3, Via Campi 213/A, 41125 Modena, Italy}

\begin{document}
\maketitle

\begin{abstract}
Being a source of clean and renewable energy, the possibility to convert solar radiation in electric current with high efficiency is one of the most important topics of modern scientific research. 
Currently the exploitation of interaction between nanocrystals seems to be a promising route to foster the establishment of third generation photovoltaics. 
Here we adopt a fully ab-initio scheme to estimate the role of nanoparticle interplay on the carrier multiplication dynamics of interacting silicon nanocrystals. 
Energy and charge transfer-based carrier multiplication events are studied as a function of nanocrystal separation showing benefits induced by the wavefunction sharing regime. We prove the relevance of these recombinative mechanisms for photovoltaic applications in the case of silicon nanocrystals arranged in dense arrays, quantifying at an atomistic scale which conditions maximize the outcome.
\end{abstract}

\begin{figure}[h]
\centering
\includegraphics[width=1.0\textwidth]{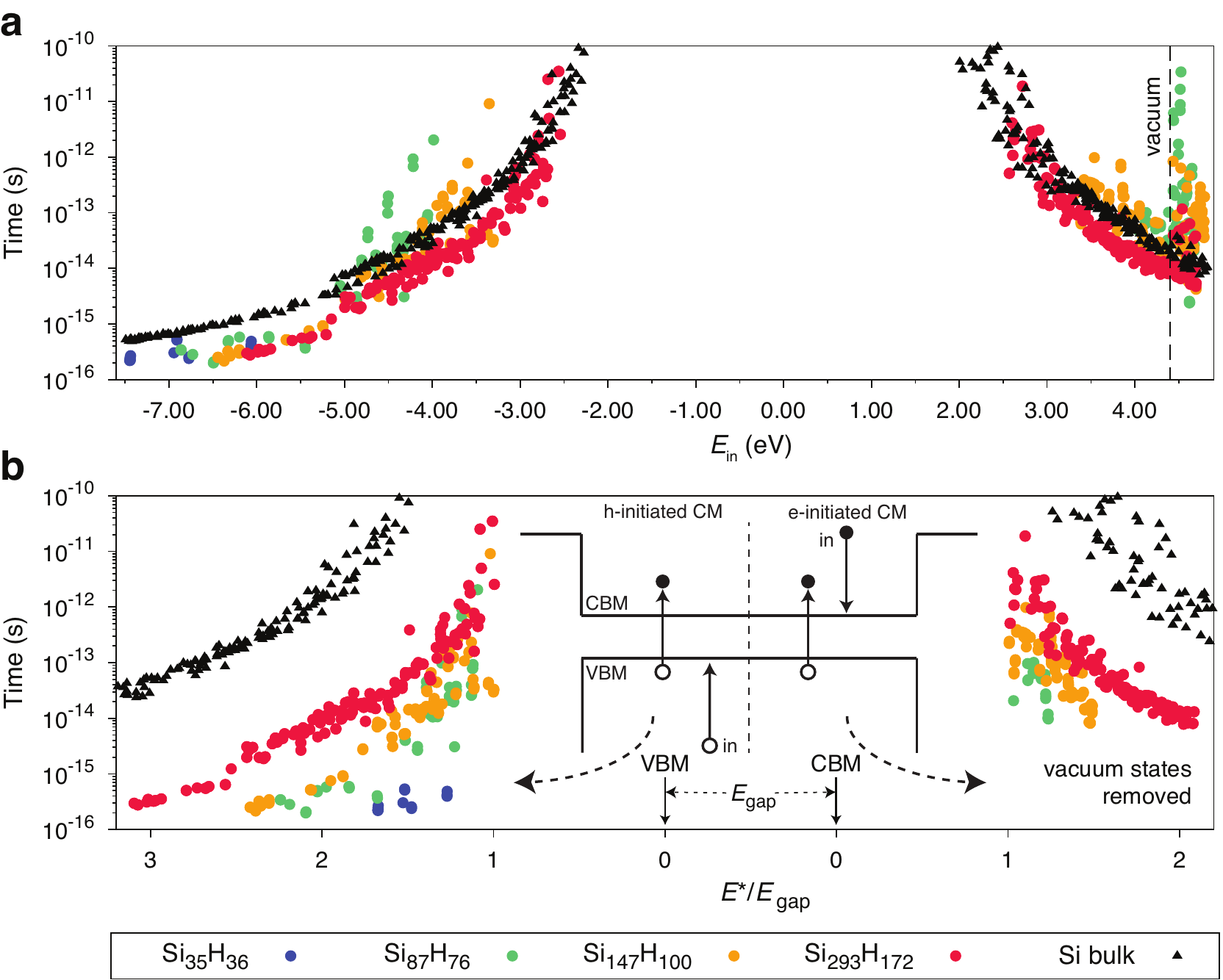}
\caption{\textbf{CM lifetimes in non-interacting Si-NCs.} Calculated CM lifetimes in isolated Si-NCs and Si-bulk.
a) CM lifetimes as a function of the energy $E_{in}$ of the carrier igniting the CM recombination (zero of energy set at half gap). Electron- (hole-) initiated CM processes are depicted on the right (left) side of the picture. The calculated vacuum level is reported with a vertical dashed line. Due to the high value of $\text{E}_{\text{gap}}$, electron-initiated CM processes in $\text{Si}_{35}\text{H}_{36}$ are energetically forbidden when unbound states are excluded. Due to the opening of the energy gap induced by quantum confinement, the smaller the NC the larger the region where one-site CM is energetically forbidden.
b) CM lifetimes as a function of the relative energy, obtained dividing the excess energy of the initial carrier by $\text{E}_{\text{gap}}$. The smaller is the NC, the stronger CM. A schematic diagram of CM decay paths (impact ionization like) is reported in the inset. VBM and CBM identify valence band maximum and conduction band minimum, respectively. The zero of the relative energy scale corresponds to the CBM (VBM), when the CM recombination is ignited by an electron (hole).}
\label{fig:isolated}
\end{figure}

\begin{figure}[h]
\centering
\includegraphics[width=0.8\textwidth]{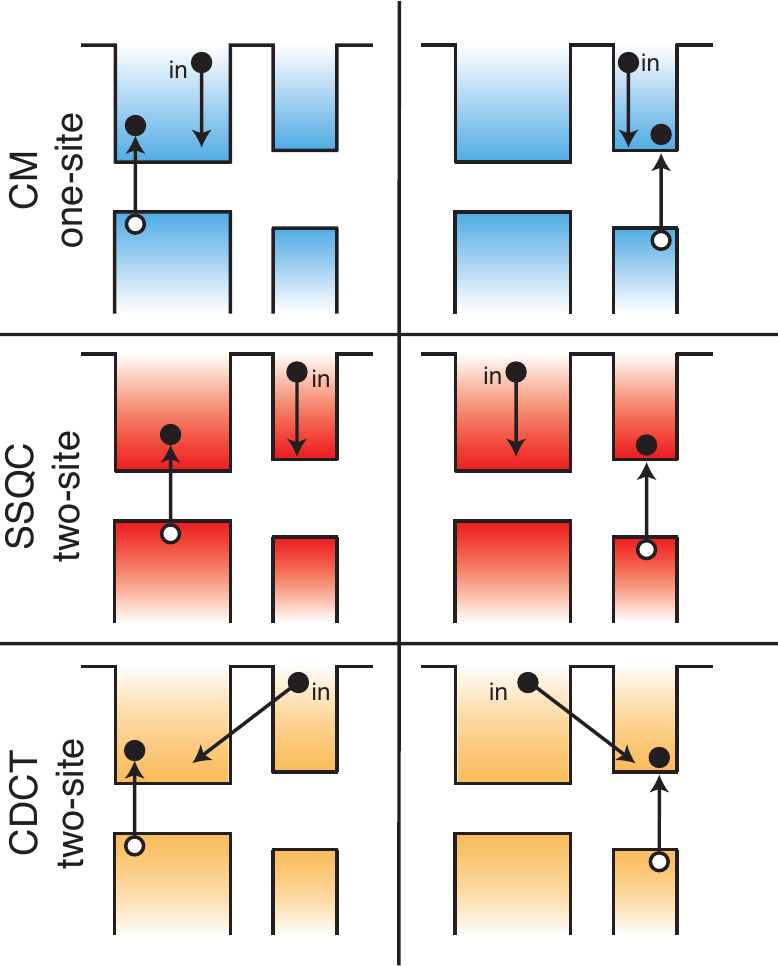}
\caption{\textbf{One- and two-site CM.} Schematics of CM events occurring in a couple of interacting NCs. The energy profile of the system of NCs is sketched (``in'' labels the carrier igniting the transition). On the contrary to one-site CM processes (in blue), which are present also in isolated NCs, two-site processes involve two nanostructures and few of them are depicted in the figure. SSQC (in red) and CDCT (in orange) processes yield a Coulomb-driven energy or charge transfer, respectively, and are discussed in the text.}
\label{fig:schemino}
\end{figure}

\begin{figure}[hbtp]
\centering
\includegraphics[width=\textwidth]{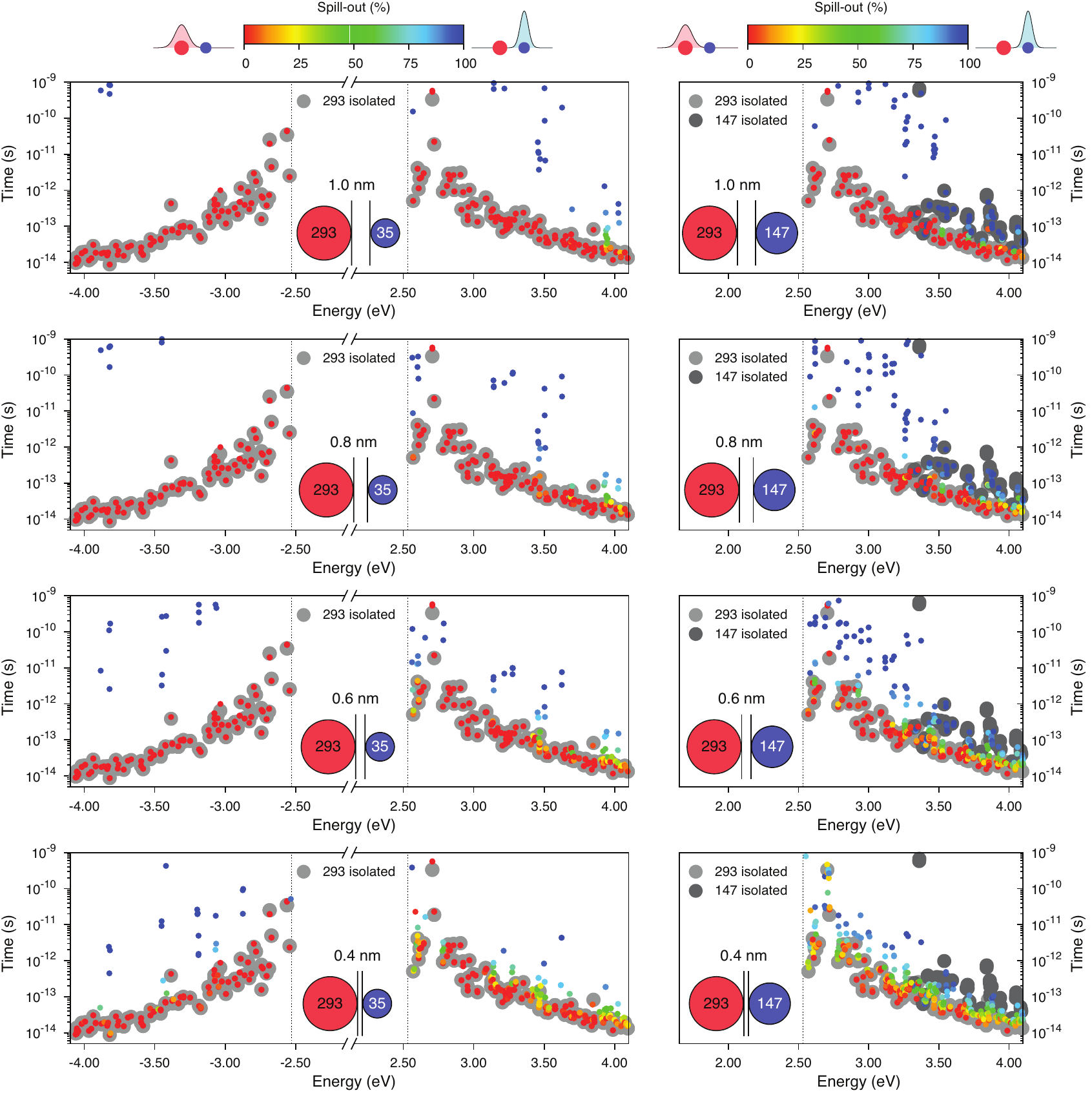}
\caption{\textbf{CM lifetimes in interacting Si-NCs.} Calculated total CM lifetimes for two sets of interacting Si-NCs. Both hole- and electron-initiated CM lifetimes for $\text{Si}_{293}\text{H}_{172}\times\text{Si}_{35}\text{H}_{36}$ are reported in the left column. Electron-initiated CM lifetimes for $\text{Si}_{293}\text{H}_{172}\times\text{Si}_{147}\text{H}_{100}$ are reported in the right column. The NC-NC separation ($1.0\,\text{nm}$, $0.8\,\text{nm}$, $0.6\,\text{nm}$ and $0.4\,\text{nm}$) corresponds to the distance between NC surfaces and is measured calculating the distance between two nearest Si atoms. A color scale supplements the results identifying where the carrier igniting the CM recombination is located. Red color (spill-out=$0\%$) is used for initial wavefunctions localized onto the large NC; blue points (spill-out=$100\%$) correspond to initial carriers localized onto the small NC. The meaning of spill-out is discussed in the text. Light and dark gray circles report the values of CM lifetimes calculated for the isolated Si-NCs, as shown in Fig.~\ref{fig:isolated}.}
\label{fig:interacting}
\end{figure}

\begin{figure}[h]
\centering
\includegraphics[width=1.0\textwidth]{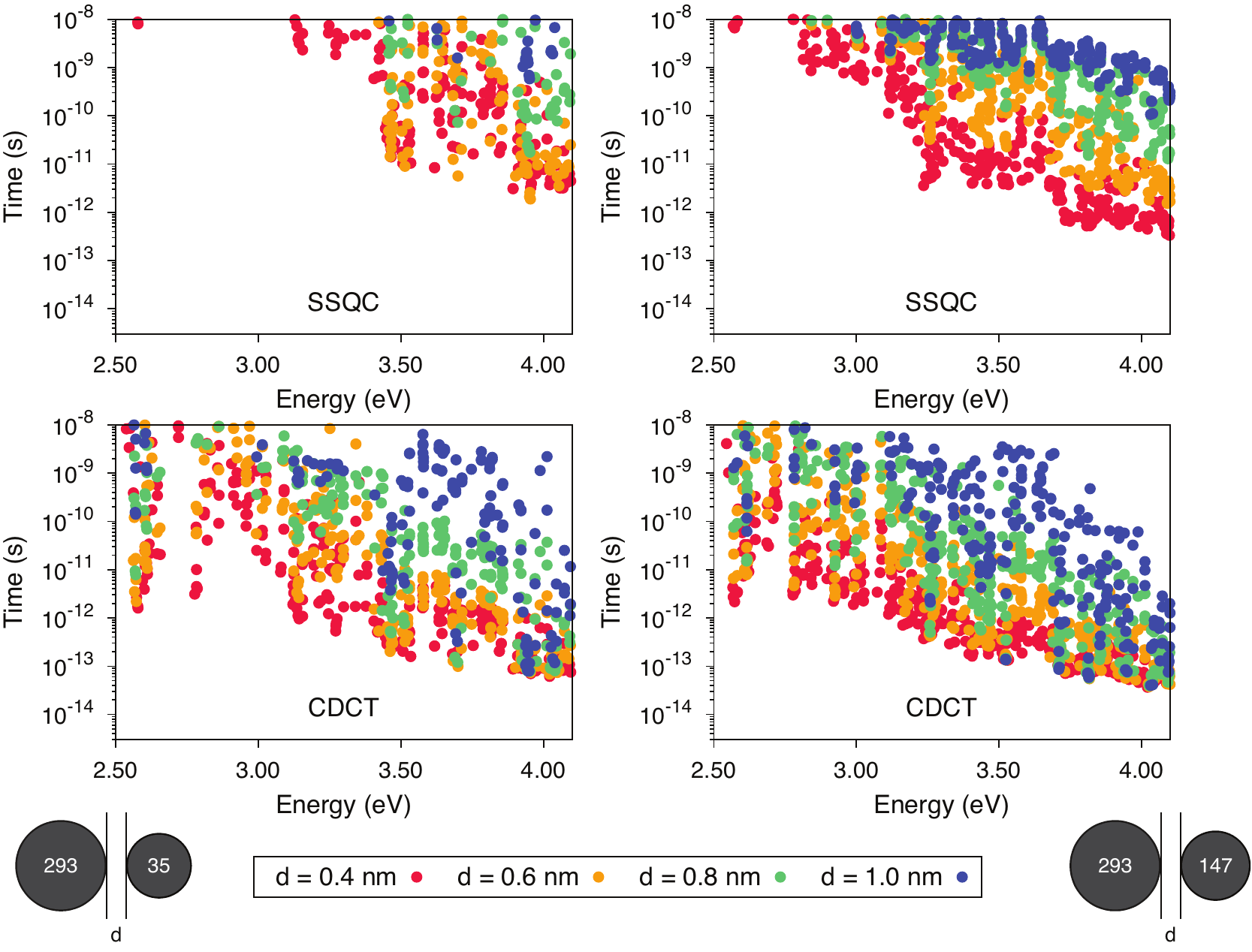}
\caption{\textbf{SSQC and CDCT lifetimes.} Calculated two-site CM lifetimes of the two considered sets of interacting Si-NCs as a function of the energy, $E_{in}$, of the carrier igniting the CM recombination. The reported electron-initiated lifetimes have been extracted from results of Fig.~\ref{fig:interacting} using weighting factors based on the spill-out computed for each involved state. As a consequence of NC-NC interaction, states of $\text{Si}_{293}\text{H}_{172}$ can delocalize into the $E_{\text{gap}}$ of the small nanostructure, yielding a reduced activation threshold with decreasing NC-NC separation. The system reported in right panels contains a bigger number of transitions due to the larger size of the NCs involved. SSQC seems to be more affected by NC-size than CDCT, becoming faster when NCs of larger size are considered.}
\label{fig:lifetimes_energy}
\end{figure}

\begin{figure}[hbtp]
\centering
\includegraphics[width=\textwidth]{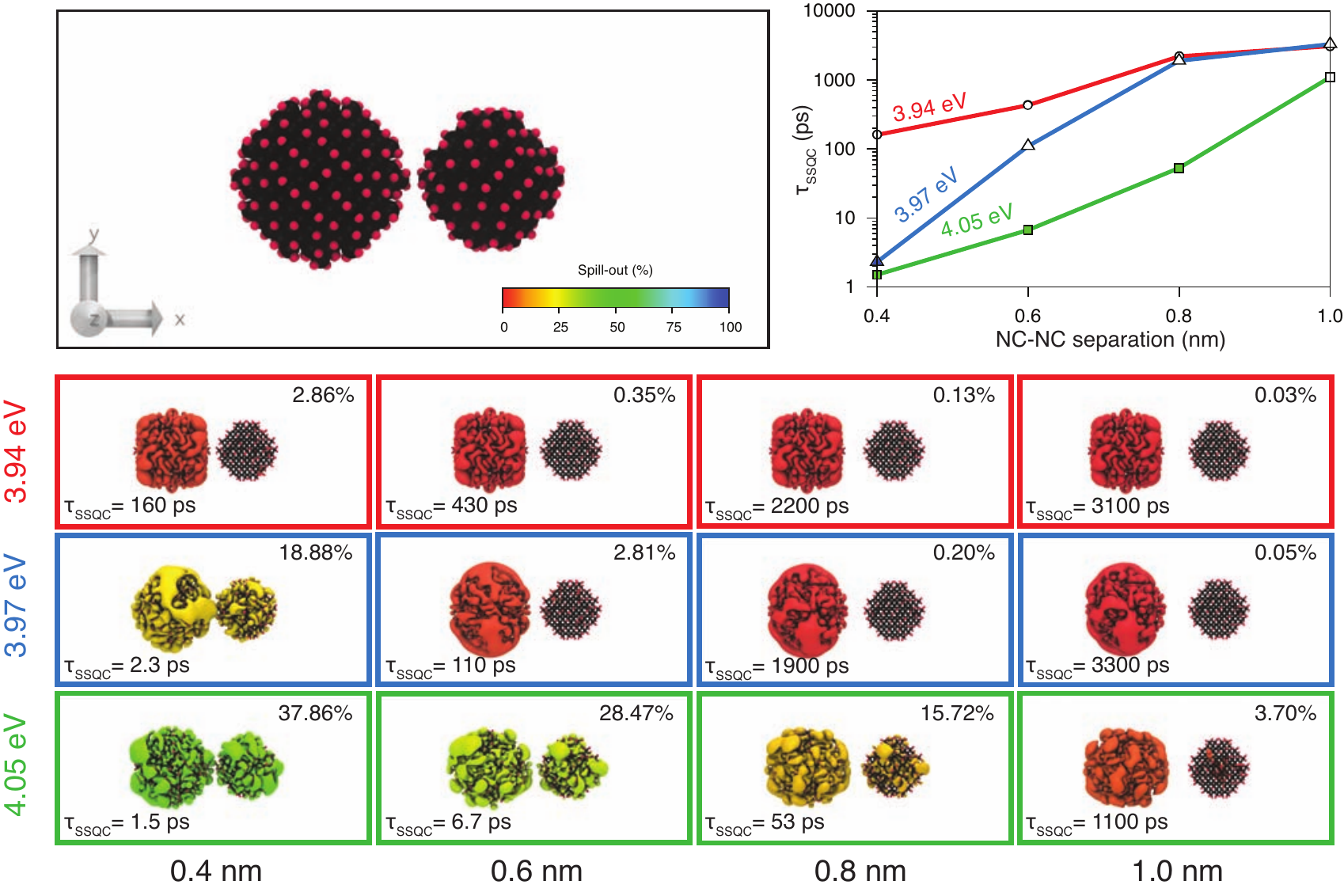} 
\caption{\textbf{Role of wavefunction delocalization.} The non-trivial dependence of SSQC lifetimes as a function of NC-NC separation is proved for the system $\text{Si}_{293}\text{H}_{172}\times\text{Si}_{147}\text{H}_{100}$. Top-left panel reports a side view of the simulation box of $(10.0\times 4.8\times 4.8)\,\text{nm}$ of size with one $\text{Si}_{293}\text{H}_{172}$ and one $\text{Si}_{147}\text{H}_{100}$ NC placed at a tunable separation and surrounded by vacuum. Bottom panels contain the isosurfaces of three KS states at several NC-NC separations. The KS states of $3.94$, $3.97$ and $4.05\,\text{eV}$ are reported with red, blue and green lines. Each isosurface has been colored according to the color scale used for the spill-out whose values are reported in the corners (together with the calculated SSQC lifetime, $\tau_{\text{SSQC}}$). The corresponding SSQC lifetimes ignited by the plotted wavefunctions are reported in the top-right panel. Here, filled triangle and square markers identify SSQC processes ignited by delocalized wavefunctions (i.e. when at least $15\%$ of the wavefunction is shared). The trends are discussed in text and highlight a general behaviour of SSQC processes whose relevance is amplified by wavefunction sharing.}
\label{fig:wfc}
\end{figure}

\begin{figure}[h]
\centering
\includegraphics[width=1.0\textwidth]{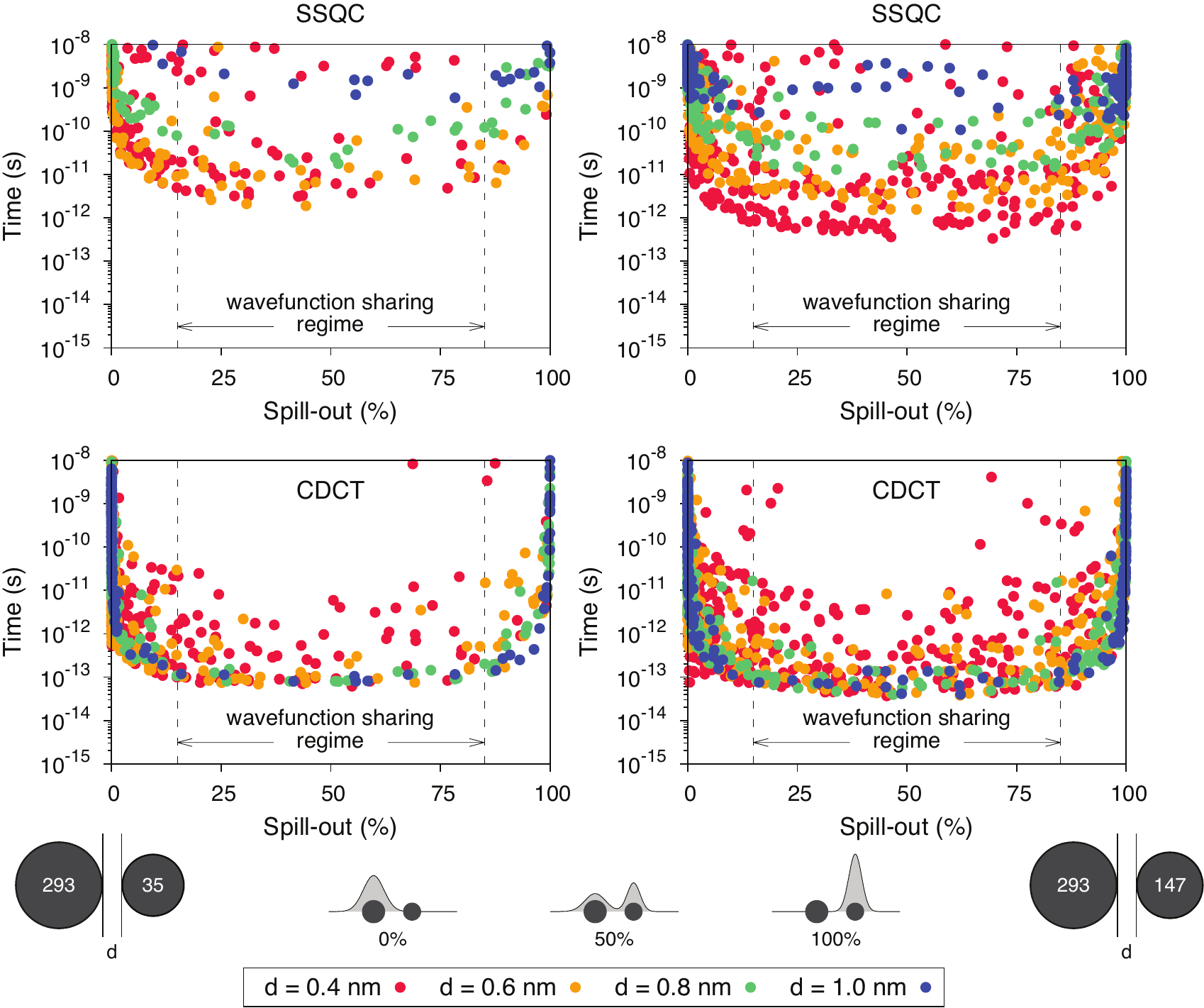}
\caption{\textbf{Wavefunction sharing regime.} Calculated two-site CM lifetimes of the two considered sets of interacting Si-NCs as a function of the spill-out of the carrier igniting the CM recombination. The reported electron-initiated lifetimes have been extracted from results of Fig.~\ref{fig:interacting} using weighting factors based on the spill-out computed for each involved state. Trends are discussed in the text. Results prove that the efficiency of both effects is boosted when both nanostructures share at least $15\%$ of the wavefunction of the initial carrier (spill-out from 15 to 85\%, defined as the wavefunction sharing regime). At $d=0.4\,\text{nm}$ slow SSQC transitions (with lifetimes of about few nanoseconds) are induced by states of the large NC that delocalize into the energy gap region of the small NC.}
\label{fig:lifetimes_spillout}
\end{figure}

A great challenge of modern science is the realization of clean, cheap and renewable energy sources. 
Among all, the possibility to convert solar energy into electric current with high efficiency is one of the most important and promising topics. 
An intense work of materials engineering is being done to increase photo-current efficiency in the visible and near-ultra violet region, through light harvesting improvement and loss factor minimization. This is typically realized by tuning the optoelectronic properties of the system or by exploiting new schemes for photovoltaic devices. 
High-energy loss factors are minimized promoting fast and non-dissipative recombination mechanisms that prevent loss thermalization processes.
In this context, carrier multiplication (CM) can be used to increase the solar cell efficiency. \\
CM is a carrier relaxation process that results in the generation of multiple electron-hole (e-h) pairs after absorption of one single photon (CM is often termed multiple exciton generation). If favoured by the quantum confinement of the electronic density, CM can be as fast as or even faster than phonon emission, extending the portion of solar spectrum converted into energy. The expected outcome is therefore an improvement of the photovoltaic performances induced by the integration of nanopatterned structures inside the solar cell device.
CM has been studied in several low dimensional nanosystems (PbSe and  PbS~\cite{ellingson_exp_PbSe_PbS,schaller_exp_PbSe,trinh_exp_PbSe,nair_exp_PbSe_PbS,schaller_seven,semonin_MEG,schaller_exp_PbSe_CdSe}, CdSe~\cite{schaller_exp_PbSe_CdSe,schaller_exp_CdSe,gachet_exp_coreshell}, PbTe~\cite{murphy_exp_PbTe}, InAs~\cite{schaller_exp_InAs} and Si~\cite{beard_exp_Si}) despite its origin and its real relevance in quantum confined structures is still under discussion~\cite{pijpers_exp_InAs,ben-lulu,mcguire_smentiscePbSe,ji_exp_PbSe,nair_exp_no_CdSe_CdTe,delerue_CMimpact,mcguire_ncvsbulk,nair_perspective}.
A particular CM scheme has been recently adopted in order to explain photoluminescence~\cite{timmerman_pssa,timmerman,timmerman_nnano} and later induced absorption~\cite{trinh} studies conducted on silicon-nanocrystals (Si-NCs) organized in dense arrays (NC-NC separation $\lesssim 1\,\text{nm}$).
In these works, Coulomb-driven energy transfer mechanisms between neighbouring Si-NCs (termed space-separated quantum cutting, SSQC) were hypothesized to generate long-lived Auger-unaffected single e-h pairs scattered among several interacting nanostructures. Distributing the excitation among several nanostructures, CM via SSQC may therefore represent one of the most suitable routes for solar cell loss factor minimization.\\
On the theoretical side, three different models have been implemented to describe CM, i.e. (i) impact ionization~\cite{califano_CdSe,delerue_impact1,franceschetti_CMimpact,allan_stativuoto,rabani_nanolett_CdSe_InAs}, (ii) coherent superpositions of single- and multi-exciton states~\cite{shabaev}, and (iii) generation of multi-excitions via virtual single-excitations (second order processes)\cite{schaller_virtual,rabani}. Despite each of them has highlighted important aspects of CM in isolated NCs, none of them has been used to investigate effects induced on CM by NC interplay due to the complexity of the problem, leaving the analysis of recombination schemes turned on by NC-NC interaction an open issue. 
A detailed study of the effects induced by the NC-NC interaction on CM dynamics and of their relevance is still missing. The mechanisms at the basis of experimental observations~\cite{timmerman,timmerman_nnano,trinh} are still unclear and the fact that SSQC can be sufficiently fast to influence CM is, to date, only a reasonable assumption.
In this context numerical simulations offer the possibility to quantify, with an accuracy that complements the experimental observations, CM events induced by NC-NC interaction and to distinguish them (here termed two-site) from the one-site CM processes. \\
In this work CM is studied, to our knowledge for the first time, adopting a fully ab-initio scheme within the density functional theory in both isolated and interacting Si-NCs. The role played by quantum confinement is, for such systems, clarified by comparing CM lifetimes calculated for isolated Si-NCs with the ones obtained for Si-bulk. 
CM events induced by NC-NC interaction (two-site CM) are quantified as a function of the separation between NCs. 
The existence of a new effect, that stems from charge-transfer processes between NCs, is proven and termed Coulomb-driven charge transfer (CDCT). A side-by-side comparison of calculated one- and two-site CM lifetimes shows the existence of a lifetime hierarchy. The impact of NC-NC interaction on CM dynamics is finally demonstrated. \\
To provide a reference for later simulations, CM in a set of isolated Si-NCs is firstly studied, simulating the ideal configuration of sparse arrays of NCs, where interactions between nanostructures can be neglected. By considering the surface dangling bonds being hydrogen-passivated, we obtain four isolated Si-NCs of different diameter: $\text{Si}_{35}\text{H}_{36}$ ($1.3\,\text{nm}$), $\text{Si}_{87}\text{H}_{76}$ ($1.6\,\text{nm}$), $\text{Si}_{147}\text{H}_{100}$ ($1.9\,\text{nm}$) and $\text{Si}_{293}\text{H}_{172}$ ($2.4\,\text{nm}$) with energy gap ($E_\text{gap}$) of $3.42$, $2.50$, $2.21$ and $1.70\,\text{eV}$, respectively. Hydrogen-passivation ensures that dangling-bond-related states are not present in the $E_\text{gap}$, a situation similar to that observed in colloidal or matrix-embedded systems. CM rates are calculated applying first order perturbation theory (Fermi's golden rule) to Kohn-Sham (KS) states~\cite{govoni_augerbulk}, thus modelling the decay of one e-h pair into two e-h pairs as the sum of two processes: one ignited by the electron (hole spectator), the other ignited by the hole (electron spectator). CM lifetimes are obtained as reciprocal of rates and considering an electron (hole), initial carrier, that decays to a negative (positive) three carrier state (termed trion~\cite{rabani_nanolett_CdSe_InAs}). The calculated one-site CM lifetimes ($\tau_{\text{one-site}}$), obtained summing over all possible final states after relaxation, are reported in Fig.~\ref{fig:isolated} as function of the energy of the initial carrier $E_{in}$ (absolute energy scale, panel a) and of the ratio $E^\star/E_{\text{gap}}$ (relative energy scale, panel b), where $E^\star$ is the excess energy of the initial carrier, i.e. measured from the respective band edge. The zero of the absolute energy scale is set at half gap ($E_{in}<0$ holes, $E_{in}>0$ electrons). 
The simulations show that CM is active when $E^\star$ exceeds $E_{\text{gap}}$ and the lifetimes monotonically decrease with $E$ from tenths of nanosecond to tenths of femtosecond (Fig.~\ref{fig:isolated}, panel a). Far from the activation threshold, CM is proved to be more efficient in Si-NCs than in Si-bulk and the lifetimes seem to be independent upon NC size resulting in an almost exact compensation between enhancement of Coulomb interaction via size reduction and decrement of density of final states. When vacuum states are counted, CM lifetimes scatter among many orders of magnitude, cease to follow the typical monotonic trend and depend upon the chosen periodic boundary conditions. The inclusion of vacuum states in CM calculations leads therefore to non-physical conclusions and hence electron-initiated CM is energetically forbidden in the smallest considered NC. 
A strong dependence of CM lifetimes upon NC size is instead observed when the relative energy scale is used. This energy scale is the most adequate to predict possible photovoltaic applications of CM~\cite{beard_ncvsbulk}, and here CM shows clear evidences of benefits induced by size reduction via quantum confinement. \\
Final states of one-site CM events are subjected to Auger recombination. This process can be responsible for re-populating high energy levels that lie above CM threshold and re-generating a configuration that can, again, decay via CM. This mechanism is called Auger recycling~\cite{navarro_augerrecycle,pitanti_augerrecycle} and its lifetime can be extracted from the effective Coulomb matrix elements so-far calculated assessing at about $1\,\text{ps}$ for the largest considered NC. An effective Coulomb matrix element can be extracted from each CM transition considering the ratio of the calculated rate and the number of final states allowed by energy conservation. \\
Starting from these considerations we extend our  analysis to the study of CM processes in dense arrays of NCs, where NC-NC interaction cannot be neglected. These configurations are simulated by placing inside the same unit cell two different Si-NCs at a tunable separation. Besides one-site CM events, NC-NC interaction opens new nonradiative CM decay channels (two-site) that stem from non-zero Coulomb matrix elements between states of different NCs. A high-energy carrier can decay to a lower energy state and transfer its energy to a nearby nanostructure where an extra e-h pair is promoted. This effect, i.e. SSQC (see Fig.~\ref{fig:schemino}), can be responsible for the generation of multiple single-excited NCs after absorption of a single photon, resulting in a distribution of the absorbed energy among separated nanostructures rather than a generation of same-site multiexciton configurations. Moreover, NC-NC interaction can also lead to nonradiative scattering processes that move one or two carriers from one NC to another, resulting in a net charge transfer with the promotion of an extra e-h pair (CDCT, see Fig.~\ref{fig:schemino}). 
Being driven by the Coulomb interaction between carriers located on separated NCs, both SSQC and CDCT are expected to occur rarely in sparse arrays of NCs and become relevant in the case of densely packed NCs where the close proximity of nanostructures boosts the two-site components of CM. In order to shed light onto the nature of these two-site CM processes we calculate, as a function of $E_{in}$, CM lifetimes for two sets of interacting Si-NCs of different size, i.e. $\text{Si}_{293}\text{H}_{172}\times\text{Si}_{35}\text{H}_{36}$ and $\text{Si}_{293}\text{H}_{172}\times\text{Si}_{147}\text{H}_{100}$, each of them being separated by a distance $d=0.4$, $0.6$, $0.8$ or $1.0\,\text{nm}$, which is close to the experimental conditions reported in Ref.~\cite{timmerman,timmerman_nnano,trinh}. Obtained results are reported in Fig.~\ref{fig:interacting} and compared with the CM lifetimes calculated for previously discussed isolated NCs (gray points). Having two NCs inside the same simulation box, wavefunctions are now free to delocalize to both nanostructures and a color scale is adopted to show the percentage (termed spill-out) of localization of the initial state: red points ($0\%$) identify CM transitions initiated by states completely localized on $\text{Si}_{293}\text{H}_{172}$, blue points ($100\%$) refer to states completely localized on the nearby NC ($\text{Si}_{35}\text{H}_{36}$ or $\text{Si}_{147}\text{H}_{100}$). Our results point out that in the system $\text{Si}_{293}\text{H}_{172}\times\text{Si}_{35}\text{H}_{36}$ and in the energy window of Fig.~\ref{fig:interacting} one-site CM is energetically forbidden inside the small NC, hence blue points are fully associated to two-site CM transitions. The other colored points can either be associated to one-site CM events occurring in the larger NC or be a mixture of one- and two-site processes. Varying $d$ from $1.0$ to $0.4\,\text{nm}$ (from top to bottom of Fig.~\ref{fig:interacting}), two-site CM lifetimes significantly decrease up to three orders of magnitude. On the contrary, CM processes ignited by states localized onto the large NC (red points) are only weakly altered by the close proximity of NCs and resemble the data obtained for the isolated $\text{Si}_{293}\text{H}_{172}$ (grey points). Modifications induced in the electronic structure by the change of the NC-NC separation therefore do not influence significantly one-site CM events.
In the system $\text{Si}_{293}\text{H}_{172}\times\text{Si}_{147}\text{H}_{100}$ (Fig.~\ref{fig:interacting}, where only CM processes ignited by excited electrons are reported), one-site CM is energetically allowed in both NCs (differently from the previous system), and Coulomb-driven charge and energy transfer processes can occur in any direction (from the small to the large NC and vice versa). At $d=1\,\text{nm}$, a group of points (spill-out $\simeq 0\%$ or $\simeq 100\%$) is superimposed to the CM lifetimes calculated for the isolated NCs. For these points, one-site CM has a predominant weight. The other transitions stem from the NC-NC interaction and represent two-site events. Also in this case, when $d$ is reduced from $1.0$ to $0.4\,\text{nm}$, two-site CM lifetimes drastically decrease up to three orders of magnitude, proving thus the strong dependence of two-site events on the NC-NC separation. \\
It is thus possible to state that in a system of interacting NCs CM recombinations involve both one- and two-site processes, the latter being strongly influenced by NC-NC separation. While it is easy to recognize the kind of CM process when wavefunctions cover just one single NC, the analysis becomes more elaborated when states are delocalized to both NCs. In order to estimate separately one-site CM, CDCT and SSQC lifetimes, we have to generalize the definition  of these effects to the case of states delocalized on the entire system. In this case each CM decay process can be split into a linear combination of one-site CM, CDCT and SSQC with the coefficients given by weighting factors that depend on the localization of the involved wavefunctions (see Supplementary Information).
Extrapolated SSQC and CDCT lifetimes ($\tau_{\text{SSQC}}$ and $\tau_{\text{CDCT}}$) are reported in Fig.~\ref{fig:lifetimes_energy} as a function of the initial carrier energy and of the NC-NC separation.
Our results  point out that  two-site CM lifetimes strongly decrease when $E_{in}$ increases or $d$ is reduced.  
Results reported in  Fig.~\ref{fig:lifetimes_energy} are fundamental because they point out the following lifetime hierarchy: $\tau_{\text{one-site}} \lesssim \tau_{\text{CDCT}} \lesssim \tau_{\text{SSQC}}$. If for instance $E_{in}=4.0 \,\text{eV}$ then $\tau_{\text{one-site}} \simeq  0.01\,\text{ps}$, $\tau_{\text{CDCT}}\simeq 0.1\,\text{ps}$ and $\tau_{\text{SSQC}}\simeq 1\,\text{ps}$ for the more realistic system $\text{Si}_{293}\text{H}_{172}\times\text{Si}_{147}\text{H}_{100}$, when NCs are placed in close proximity. %
When combined with Auger recycling, this lifetime hierarchy can be used to monitor the after-pumping excited states population, simulating thus a pump-and-probe experiment. A cyclic procedure of fast one-site CM, two-site CM and Auger recycling, plugged into a set of rate equations, is compatible with recent experimental evidences of quantum cutting in dense arrays of Si-NCs~\cite{timmerman,timmerman_nnano,trinh}. 
The predicted time-dependent population of excited states is reported and commented in the Supplementary Information, where it is shown that a double number of e-h pairs with respect to the number of absorbed above-CM-threshold photons appears directly after pumping (in low fluence conditions). One-site CM is responsible for probing a double number of e-h pairs soon after short delay times ($\tau_{\text{one-site}}\simeq 0.01\,\text{ps}$), while SSQC keeps this number long-lived and subjected to radiative-decay only. 
Noticeably, exciton recycling mechanisms were already hypothesized in order to interpret energy transfer mechanisms in rare earth (Er$^{3+}$) coupled Si-NCs~\cite{navarro_augerrecycle,pitanti_augerrecycle}.\\
Let us monitor the two-site CM dependence upon NC proximity. In order to find the conditions that maximize quantum cutting we firstly select three different initial states close in energy ($E_{in}=3.94$, $3.97$ and $4.05\,\text{eV}$, respectively) that show a different localization when NC-NC separation is reduced, and we report the SSQC lifetimes for each of them as a function of $d$ and of the spill-out variable (see Fig.~\ref{fig:wfc}).
A strong localization onto the large NC is observed for all three initial states when $d=1\,\text{nm}$ and the SSQC lifetimes assess at few nanoseconds. Whenever the initial state delocalizes to both NCs, a SSQC lifetime drop is observed showing that SSQC becomes strikingly more efficient in this regime. This occurs at $d\leq 0.8\,\text{nm}$ for the transition ignited by the state with $E_{in}=4.05\,\text{eV}$ and at $d\leq 0.4\,\text{nm}$ for the state with $E_{in}=3.97\,\text{eV}$. Such behaviour is not observed when $E_{in}=3.94\,\text{eV}$ due to the lack of delocalization and hence SSQC remains noticeably the slowest. It is evident therefore that SSQC lifetimes show a non-trivial dependence upon the NC-NC separation and strongly depend on the localization of the initial state; they undergo a dramatic reduction when the initial carrier wavefunction delocalizes over the entire system. In other words, SSQC processes become fast when they cease to be simple e-h generation mechanisms driven by energy transfer processes between separated and independent NCs and start to probe the compound system of two NCs as unique from the quantum point of view. To prove that this is a general trend of SSQC, not limited to the three selected states, and to extend the same considerations to CDCT, we report, in Fig.~\ref{fig:lifetimes_spillout}, the calculated SSQC and CDCT lifetimes as a function of the spill-out for all the NC-NC separations taken into account. As previously noted,  two-site lifetimes are large when the spill-out is either $0\%$ or $100\%$, i.e. when the initial wavefunction is strongly localized onto one single nanostructure. On the contrary, even a small delocalization of the initial wavefunction on both nanostructures is able to push up two-site Coulomb matrix elements. We observe changes up to two-three orders of magnitude in both SSQC and CDCT lifetimes when the initial state ceases to be completely localized onto one NC and at least $15\%$ of the wavefunction is shared by the two NCs. The maximum efficient two-site events are recorded when the initial carrier wavefunction extends to both NCs and the spill-out parameter ranges about from $15\%$ to $85\%$, which defines the wavefunction sharing regime. These results point out that pictures where energy and charge transfer mechanisms occur among independent NCs with wavefunctions unaffected by NC proximity (F\"orster-like processes) do not provide a good description of CM recombinations induced by NC-NC interaction (see Supplementary Information). Only a full quantum picture of the system, considered as a whole, properly describes the CM dynamics of two interacting NCs in the considered range of separations (which is the one worthwhile for photovoltaic applications). Favouring the creation of orbitals shared by many NCs, the embedding matrix~\cite{aerts_PbSe} or the presence of several interacting NCs (typical condition of three-dimensional realistic systems) is expected to amplify the relevance of SSQC and CDCT in dense arrays of Si-NCs.

In conclusion, we calculated the typical lifetimes of CM processes occurring in both isolated and interacting Si-NCs. Simulations of CM lifetimes in isolated Si-NCs showed that CM benefits from size reduction via quantum confinement. When the interaction between NCs was turned on, CM recombinations were separated into one-site and two-site CM processes, depending on whether the transitions involved only one single NC or couples of NCs. Both Coulomb-driven energy and charge transfer mechanisms were studied as a function of the NC-NC separation generalizing the common view of two separated interacting NCs to a unique quantum confined structure where quantum states can spread among both nanostructures. In particular wavefunction delocalization gave to the two-site CM dynamics a non-trivial dependence upon NC-NC separation. For the first time both SSQC and CDCT lifetimes were estimated showing optimal efficiencies in the wavefunction sharing regime that can be favoured by the presence of several interacting NCs (typical of dense arrays of NCs). 
On the basis of ab-initio simulations, the presence of a cyclic procedure of fast one-site CM, two-site CM and Auger recycling mechanisms was proven to be compatible with the recent experimental observations of quantum cutting in Si-NCs based on photoluminescence~\cite{timmerman,timmerman_nnano} and induced absorption experiments~\cite{trinh}. 
Thanks to the ability of generating long-lived single e-h pairs after distribution of the excitation among interacting nanostructures, SSQC is expected to have great impact on solar cell devices based on Si-NCs as long as the NCs are arranged in dense arrays. 
Tailoring the outcome by tuning NCs proximity is hence possible to use NC-NC interaction to overcome the current solar cell limitations opening a new route to foster the establishment of third generation photovoltaics.

\vspace{0.5cm}

\section*{Methods}

Electronic structures have been calculated from first-principles within the density functional theory in the local density approximation, adopting a pseudopotential supercell approach in reciprocal space~\cite{giannozzi}. Quasiparticle corrections have been applied to the electronic structure of Si bulk~\cite{marini}. On the contrary to the elsewhere used real-space methods, our methodology offers a natural description of the electronic and screening properties of both k-dispersive and low-dimensional systems, treating them on an equal footing. CM lifetimes are obtained using first order perturbation theory (impact ionization scheme) calculating the screened Coulomb matrix elements between KS states, for both isolated and interacting NCs. The dielectric screening is obtained solving the Dyson's equation for the polarizability in the random phase approximation. 
For couples of interacting NCs of different size, localization of the wavefunctions is measured by introducing the spill-out that represents the probability of having the state localized onto the small NC. 
A general definition of one-site and two-site CM lifetimes that takes into account wavefunction delocalization is reported in the Supplementary Information.


\bibliography{manuscript_GOVONI}

\section*{Acknowledgements}
We acknowledge the Super-Computing Interuniversity Consortium CINECA for the availability of support and high performance computing resources under the Italian Super-Computing Resource Allocation (ISCRA) initiative. We acknowledge the European Community's Seventh Framework Programme (FP7/2007-2013) under Grant Agreement No. 245977. We thank G. Cantele and F. Iori for fruitful discussions.

\section*{Author contributions}
M.G., I.M and S.O. conceived the project. M.G. and I.M. designed and performed simulations with a code developed by M.G., and co-wrote the manuscript. I.M. and S.O. supervised the project. All authors discussed the results and implications and commented on the manuscript at all stages.

\section*{Additional information}
The authors declare no competing financial interests. Supplementary Information accompanies this paper. Correspondence and requests for materials should be addressed to M.G. and I.M.

\end{document}